\documentclass[useAMS,usenatbib,usegraphicx]{mn2e}
\usepackage{epsf}

\title[The Quadruply Lensed Quasar SDSS J1330+1810]
{Mass models and environment of the new quadruply lensed quasar 
SDSS J1330+1810\thanks{This paper includes data gathered
with the 6.5 meter Magellan Telescopes located at Las Campanas
Observatory, Chile. Based on observations obtained with the Apache
Point Observatory 3.5-meter telescope, which is owned and operated by
the Astrophysical Research Consortium. Use of the UH 2.2-m telescope
for the observations is supported by NAOJ. }}

\author[M. Oguri et al.]
{Masamune Oguri,$^1$\thanks{E-mail: oguri@slac.stanford.edu} 
Naohisa Inada,$^2$
Jeffrey A. Blackburne,$^3$ 
Min-Su Shin,$^4$ \newauthor
Issha Kayo,$^{5,6}$ 
Michael A. Strauss,$^4$ 
Donald P. Schneider,$^7$ and
Donald G. York$^{8,9}$\\
$^1$Kavli Institute for Particle Astrophysics and Cosmology, 
Stanford University, 2575 Sand Hill Rd., Menlo Park, CA
94025, USA.\\ 
$^2$Cosmic Radiation Laboratory, RIKEN, 2-1 Hirosawa, Wako, Saitama 
 351-0198, Japan.\\ 
$^3$Department of Physics, Massachusetts  
Institute of Technology and Kavli Institute for Astrophysics and  
Space Research, \\
70 Vassar St., Cambridge, MA, 02139, USA.\\
$^4$Princeton University Observatory, Peyton Hall,
 Princeton, NJ 08544, USA.\\
$^5$Institute for the Physics and Mathematics of the Universe,
 University of Tokyo, 5-1-5 Kashiwanoha, Chiba 277-8582, Japan.\\
$^6$Department of Physics and Astrophysics, Nagoya
University, Chikusa-ku, Nagoya 464-8602, Japan.\\
$^7$Department of Astronomy and Astrophysics, 
Pennsylvania State University, 525 Davey Laboratory, 
University Park, PA 16802, USA.\\
$^8$Department of Astronomy and Astrophysics, The University 
 of Chicago, 5640 South Ellis Avenue, Chicago, IL 60637, USA.\\
$^9$Enrico Fermi Institute, The University of Chicago,
5640 South Ellis Avenue, Chicago, IL 60637, USA.
} 

\begin{document}

\date{\today}

\voffset- .5in

\pagerange{\pageref{firstpage}--\pageref{lastpage}} \pubyear{}

\maketitle

\label{firstpage}

\begin{abstract}
We present the discovery of a new quadruply lensed quasar. The lens 
system, SDSS J1330+1810 at $z_s=1.393$, was identified as a lens
candidate from the spectroscopic sample of the Sloan Digital Sky 
Survey. Optical and near-infrared images clearly show four
quasar images with a maximum image separation of $1\farcs76$, as well
as a bright lensing galaxy. We measure a redshift of the lensing
galaxy of $z_l=0.373$ from absorption features in the spectrum. We
find a foreground group of galaxies at $z=0.31$ centred $\sim 120''$
southwest of the lens system. Simple mass models fit the data
quite well, including the flux ratios between images, although the
lens galaxy appears to be $\sim 1$~mag brighter than expected by the
Faber-Jackson relation. Our mass modelling suggests that shear from
nearby structure is affecting the lens potential.   
\end{abstract}


\section{Introduction}
\label{sec:intro}

\begin{figure*}
\begin{center}
 \includegraphics[width=0.75\hsize]{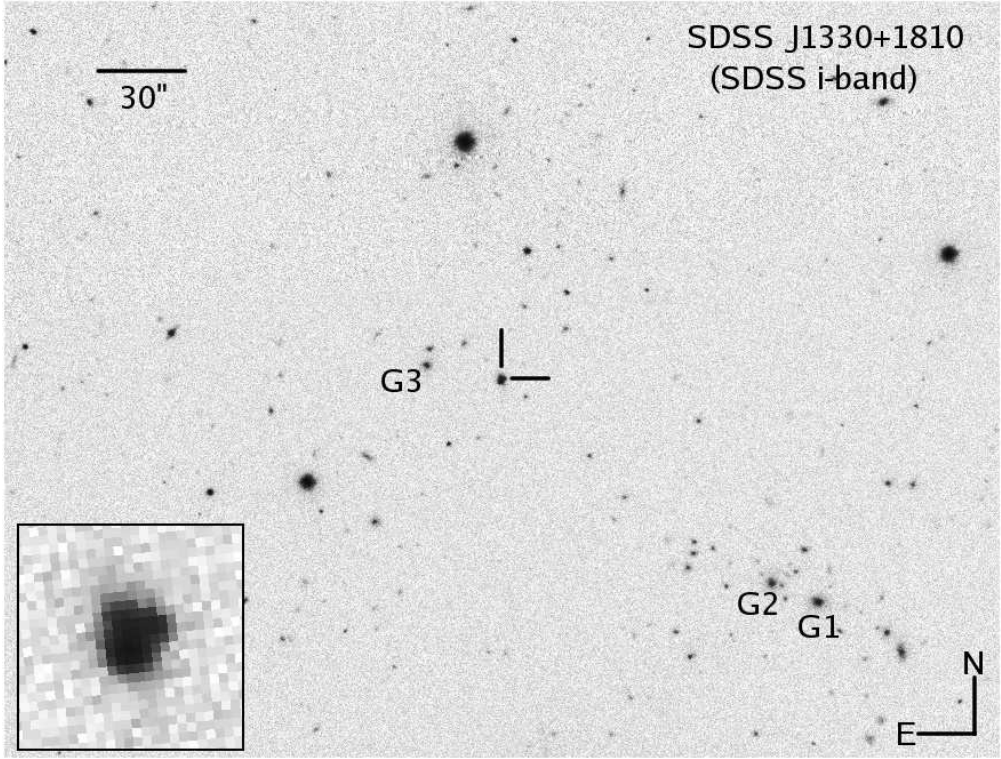}
\end{center}
\caption{The SDSS $i$-band image of the SDSS~J1330+1810 field. The
  pixel scale of the image is $0\farcs396$~${\rm pixel^{-1}}$, and the
  seeing was $1\farcs0$. The inset at lower left shows an  expanded
  view of the lens system. Several bright galaxies in the field are
  indicated by G1-G3. The $i$-band Petrosian magnitudes of these
  galaxies (without correcting for Galactic extinction) are $17.72$
  (G1), $18.20$ (G2), and $18.59$ (G3).  
\label{fig:fc1330}}
\end{figure*}

\begin{figure}
\begin{center}
 \includegraphics[width=0.95\hsize]{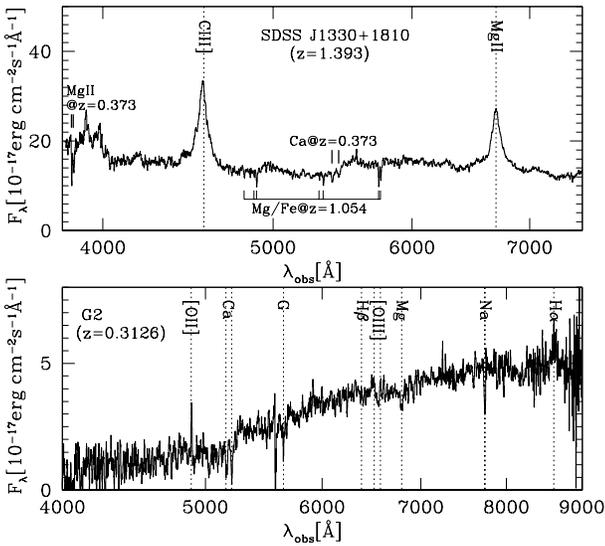}
\end{center}
\caption{The SDSS spectra of SDSS~J1330+1810 ({\it upper}) and the 
nearby galaxy G2 ({\it lower}). See Figure \ref{fig:fc1330} for
the relative positions. The spectral resolution is $R\sim 2000$, but 
the spectra are smoothed with a 5-pixel boxcar to suppress
noise. Thick vertical line segments in the upper panel mark MgII and
Ca absorption lines, which suggest the lens redshift of $z_l=0.373$. 
A Mg/Fe absorption system at $z=1.054$ is also shown by thin lines. 
The feature at $\sim 5600${\AA} in the spectrum of G2 is due
to poor subtraction of a strong sky emission line ($5575${\AA}). 
\label{fig:spec1330}}
\end{figure}

Thus far about 100 gravitationally lensed quasars are known, of
which $\sim30$ are quadruple (four-image) lenses. The number ratio 
of quadruple lenses to double (two-image) lenses contains information 
on both the shapes of lensing galaxies and the luminosity function of
source quasars \citep{rusin01,chae03,huterer05,oguri07b,madelbaum08}. 
In addition, quadruple lenses allow more detailed mass modelling of
individual lenses. For instance, the larger number of images provides
more constraints on the lens potential, which is essential in probing
the effects of external perturbations on primary lenses 
\citep{keeton97} and constraining the Hubble constant from time delay 
measurements \citep[e.g.,][]{suyu06}. Magnifications of merging image
pairs in quadruple lenses satisfy distinct relations if the lens
potential is smooth, but small-scale structures near the image can
violate the relations. Thus flux ratios of quadruple lens images serve
as unique probes of substructure or microlensing in lens galaxies 
\citep{mao98,metcalf01,chiba02,dalal02,schechter02}.   

In this paper, we present the discovery of a new gravitationally
lensed quasar with four lensed images, SDSS~J133018.65+181032.1
(SDSS~J1330+1810). It was discovered as part of the Sloan Digital Sky
Survey Quasar Lens Search \citep[SQLS;][]{oguri06,oguri08,inada08},
which takes advantage of the large spectroscopic quasar catalog
\citep[see][]{schneider07} of the Sloan Digital Sky Survey
\citep[SDSS;][]{york00} to locate new lensed quasars. We place
particular emphasis on mass modeling and investigation of the
structure around the lens. 

The outline of this paper is as follows. We describe the SDSS and
follow-up data in Sections \ref{sec:sdss} and \ref{sec:followup},
respectively. The environment of the lens is discussed in Section
\ref{sec:env}. Section \ref{sec:model} is devoted to mass modelling.
Our results are summarised in Section \ref{sec:sum}. 
Throughout the paper, we adopt the standard Lambda-dominated flat
universe cosmology with matter density $\Omega_M=0.26$ and the Hubble
constant $h=0.72$ \citep{dunkely08}.

\section{The SDSS Data}
\label{sec:sdss}

\begin{figure*}
\begin{center}
 \includegraphics[width=0.8\hsize]{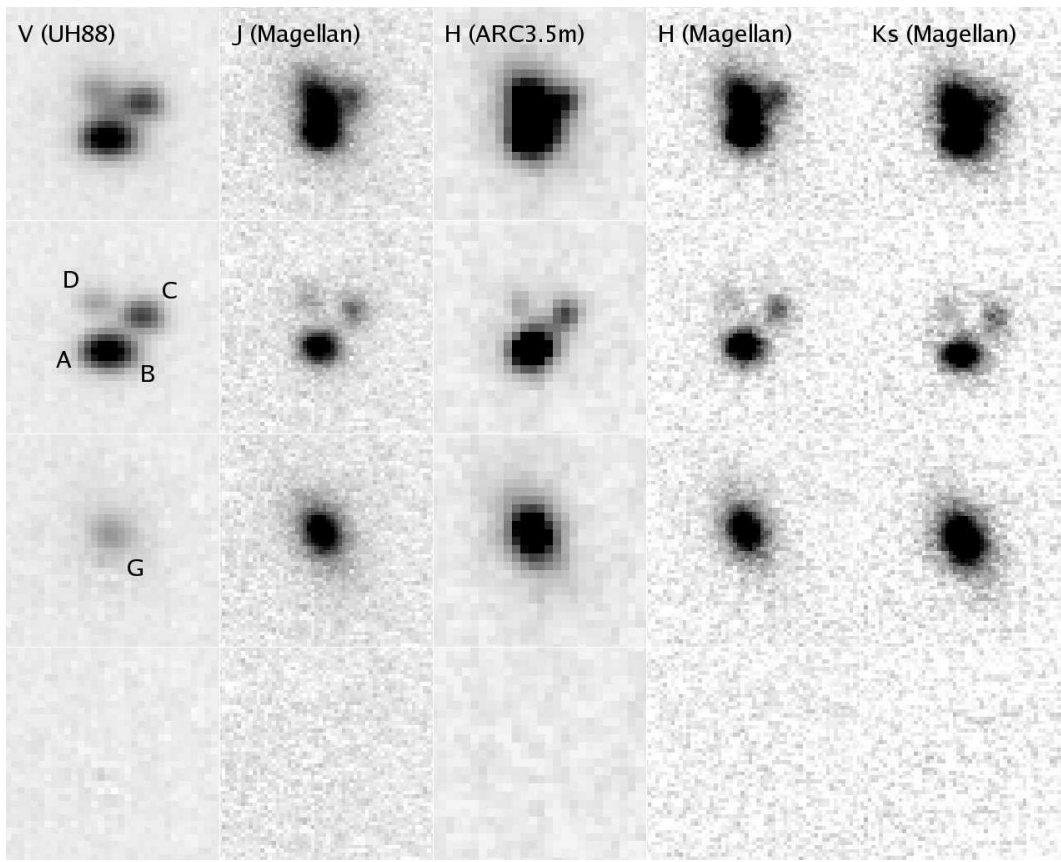}
\end{center}
\caption{Images of SDSS J1330+1810. All the panels are
  $7''\times7''$ in size. North is up and East is left. The top row
  shows original follow-up images, the second row displays images
  after subtracting the lensing galaxy G1, the third row contains
  images after subtracting 4 PSFs (quasar images), and the bottom row
  shows the  results after subtracting the galaxy and all four quasar
  components. Images for each band are displayed with the same
  gray-level stretch. 
\label{fig:image1330}}
\end{figure*}

\begin{table*}
 \caption{Relative astrometry and photometry from the follow-up images
   (see Figure \ref{fig:image1330}). The positive directions of $x$
   and $y$ are North and West, respectively. The J2000 coordinates of 
   component A are (RA, Dec) = (202.57784, 18.17562). The positional
   errors are estimated from the scatter between the five follow-up
   images.The errors on the magnitudes are
   statistical errors only and do not include systematic errors coming
   from uncertainties of PSFs and the galaxy profile. The magnitudes
   have not been corrected for Galactic extinction. The magnitudes of
   galaxy G refer to the total magnitudes. 
\label{table:sdss1330}} 
 \begin{tabular}{@{}cccccccc}
  \hline
   Name & $x$ & $y$ &  $V$ (UH88) & 
   $J$ (Magellan) & $H$ (ARC3.5m) & $H$ (Magellan) &
   $K_s$ (Magellan) \\
   & [arcsec] & [arcsec] & [mag] & [mag] & [mag] & [mag] & [mag] \\
   \hline
  A & $\equiv 0$ & $\equiv 0$ & $19.02\pm0.02$ &  $18.36\pm0.05$ & 
  $17.32\pm0.05$ & $17.40\pm0.07$ & $17.06\pm0.04$ \\
  B & $0.42\pm0.03$ & $-0.01\pm 0.03$ & $19.72\pm0.03$ &
  $18.68\pm0.07$ &  $17.54\pm0.06$ & $17.73\pm0.09$ & $17.30\pm0.05$
  \\
C &  $1.30\pm0.03$ &  $1.19\pm0.03$ & $19.89\pm0.01$ &  $19.12\pm0.02$ 
  &  $18.18\pm0.02$ & $18.24\pm0.02$ & $17.90\pm0.03$\\ 
D & $-0.24\pm 0.04$ & $1.58\pm0.04$ & $21.45\pm0.04$ &  $19.83\pm0.05$ 
  &  $19.36\pm0.12$ &  $19.13\pm0.05$ & $18.62\pm0.06$\\ 
G &  $0.24\pm0.03$ & $0.97\pm0.03$ & $19.48\pm0.04$ & $16.77\pm0.01$ 
  &  $16.00\pm0.01$ & $16.08\pm0.02$ & $15.26\pm0.01$\\ 
 \hline
 \end{tabular}
\end{table*}

The lens system SDSS~J1330+1810 was first identified in the data of
the SDSS-II \citep{adelman08}. The SDSS-II is a survey to map 10,000 
square degrees of the northern sky with a dedicated wide-field
2.5-meter telescope \citep{gunn06} at the Apache Point Observatory in
New Mexico, USA. It consists of a photometric survey
\citep{gunn98} with five broad-band optical filters
\citep{fukugita96} and a spectroscopic survey  of quasars and
galaxies selected by a series of target selection algorithms;
\citet{richards02} present the SDSS quasar selection technique, and
\citet{blanton03} describe the generation of the final SDSS
spectroscopic targets.  The homogeneity and good quality of the 
data, with an astrometric accuracy better than $0\farcs1$ rms
per coordinate \citep{pier03} and a photometric zeropoint accuracy
better than 0.02 magnitude over the entire survey area
\citep{hogg01,smith02,ivezic04,tucker06,padmanabhan08}, are essential
for various statistical studies. 

The SQLS identifies gravitationally lensed quasar candidates using
a well-defined algorithm \citep{oguri06} applied to the spectroscopic
SDSS quasars. The algorithm has two parts: morphological selection,
which identifies quasars that are poorly fitted by the Point Spread
Function (PSF), and colour selection, which examines objects near
each spectroscopic quasar and selects those with colours similar to
the quasar as lens candidates. These two selections are designed to
locate small- ($\sim 1''$) and large-separation ($\ga 3''$) lensed
quasars, respectively. The SQLS has already discovered $>20$ new
gravitationally lensed quasars from the SDSS data using this
algorithm, including both double and quadruple lenses
\citep[e.g.,][and references therein]{kayo07}.  

The gravitational lens SDSS~J1330+1810 was selected as a lens
candidate by the morphological selection algorithm.\footnote{The SDSS
photometric designation of SDSS~J1330+1810 is 5308/40/1/73/88
(run/rerun/camcol/filed/ID), and the spectroscopic designation is
2641/255/54230 (plate/fiber/MJD).}
Figure \ref{fig:fc1330}
shows the SDSS $i$-band image of the system (seeing of
$1\farcs0$). The enlarged image clearly indicates that the system,
which is classified as a quasar at $z=1.393$ from the SDSS spectrum
(see Figure \ref{fig:spec1330}), is not a point source but consists of
multiple components. The morphology of the quasar in the SDSS image is
similar to that of SDSS~J0924+0219 \citep{inada03}, suggesting that
this is likely to be a fold-type quadruple lens.  
We find no bright radio \citep[FIRST;][]{becker95} or X-ray
\citep[RASS;][]{voges99} source in the vicinity of SDSS~J1330+1810. 
In the SDSS spectrum
we find absorption lines at $\sim 3850${\AA} and $\sim 5400${\AA},
consistent with MgII and Ca lines due to a galaxy at $z=0.373$. We
interpret these features as absorptions by the lensing galaxy. In
addition a Mg/Fe absorption system at $z=1.054$ is seen in the
spectrum. Since its redshift is quite close to the source redshift,
the effect of the absorber at $z=1.054$ on the lens potential is
expected to be small. 

The SDSS image shows a concentration of red galaxies $\sim 120''$
southwest of SDSS~J1330+1810 (see Figure \ref{fig:spec1330}, around G1
and G2). One of the galaxies, G2, was targeted by the luminous red
galaxy program \citep{eisenstein01} and has been observed
spectroscopically in the SDSS (Figure \ref{fig:spec1330}), with a
redshift of $z=0.3126$. The presence of the [OII] emission line
(rest-frame equivalent width of $-9${\AA}) suggests ongoing star
formation activity in galaxy G2. We explore the possible group further
in Section~\ref{sec:env}.    

\section{Followup Observations}
\label{sec:followup}

\subsection{Optical and Near-Infrared Imaging}
\label{sec:image}

Near-infrared ($JHK_s$-bands) images were taken using Persson's
Auxilliary Nasmyth Infrared Camera \citep[PANIC;][]{martini04} at the
6.5-meter Magellan I (Walter Baade) telescope on 2007 July 5. The
total exposure time was  405~sec in $J$, 540~sec in $H$, and 360~sec
in $K_s$. The seeing was $0\farcs 74-0\farcs82$. We also obtained
$H$-band images with the Near-Infrared Camera and Fabry-Perot
Spectrometer (NICFPS) at the Astrophysical Research Consortium
3.5-meter (ARC3.5m) Telescope on 2007 June 1. The total exposure time
was 1200~sec and the seeing was $0\farcs88$. Optical
imaging was conducted with the Tektronix 2048x2048 CCD
camera at the University of Hawaii 2.2-meter (UH88) telescope. A
500~sec image in $V$-band was taken on 2008 
March 6, under $0\farcs81$ seeing. All the data were reduced
using standard IRAF tasks. The zero-point magnitudes of the
infrared images were estimated using Two Micron All Sky Survey
(2MASS) data \citep{skrutskie06}, whereas the UH88 image was
calibrated by the standard star PG0918+029 \citep{landolt92}.
 
The images shown in Figure \ref{fig:image1330} confirm that
the system is indeed a typical fold-type quadruple lens, with two
merging images and the other two images on the other side of the lens.
For definiteness, we fit the images using GALFIT \citep{peng02}. The
assumed model consists of four PSFs and a lens galaxy modelled by a
Sersic profile with the convolution of the PSF. We adopt nearby stars
as PSF templates. We first left the Sersic index $n$ as a free parameter 
and found the best-fit value to be $n\approx 3.4$, which is close to
the canonical value for early type galaxies, $n=4$. Thus in what
follows we fix the Sersic index to $n=4$. We find that this model fits
the data quite well. The subtracted images shown in Figure
\ref{fig:image1330} show virtually no residuals. Table
\ref{table:sdss1330} summarises the relative astrometry and photometry
from the fitting. Following  convention, we name four quasar
components A-D, in decreasing order of their brightnesses, and we name
the lensing galaxy G. The relative positions of the four components
agree well among the high angular resolution images, with a scatter of
$\sim0\farcs03$.  The maximum separation between images is
$1\farcs76$. For the NIR images, the lensing galaxy is modelled well
by the Sersic profile ($n=4$) with scale radius $R_e\sim 0\farcs
7-0\farcs9$, ellipticity $e\sim 0.57$, and position angle (East of
North) $\theta_e\sim 24^\circ$. The lensing galaxy has somewhat
different best-fit scale radius and ellipticity in the UH88 $V$-band
image, $R_e\sim 1\farcs 42$ and $e\sim0.28$.  The larger $R_e$ in
$V$-band is in fact consistent with observations of elliptical
galaxies \citep[e.g.,][]{pahre99}

\begin{figure}
\begin{center}
 \includegraphics[width=0.9\hsize]{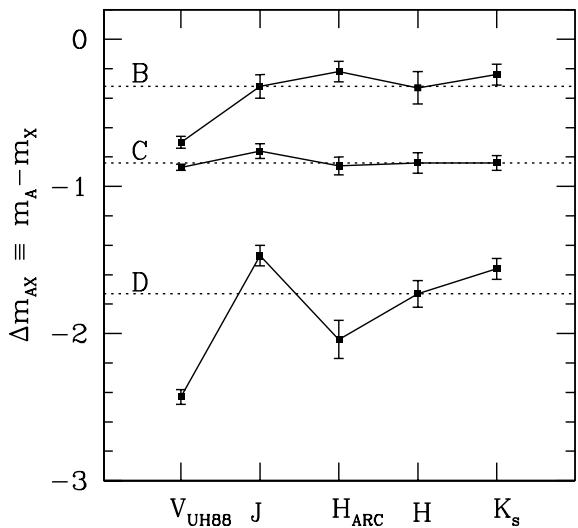}
\end{center}
\caption{Flux ratios of quasar images from the follow-up images (see
  also Figure \ref{fig:image1330} and Table \ref{table:sdss1330}). We
  plot magnitude differences between images B-D and the brightest
  image A, $\Delta m \equiv m_A-m_X$ ($X=B-D$). Dotted horizontal
  lines indicate median values of individual ratios, which we adopt
  for mass modelling. \label{fig:fratio}} 
\end{figure}

In Figure \ref{fig:fratio}, we compare the flux ratios of the quasar
images derived from the five follow-up images. They are broadly
consistent between different wavelengths, which 
further supports the lensing interpretation. However, there are some
noticeable differences in flux ratios, in particular those in the UH88 
$V$-band. While additional systematic errors associated with image
fitting might account for the differences, a possible physical
interpretation is differential dust extinction which is commonly seen
in lensed quasar systems \citep[e.g.,][]{falco99}; this helps to
explain the different $V$-band flux ratios from those in NIR 
images, because the effect of dust is much more pronounced in $V$-band
than in NIR. The $V$-band image was taken $8-9$ months after the
observations in near-infrared, and thus time variability (intrinsic
and/or due to microlensing) of the quasar images might also
contribute to the differences. In particular, the fact that image B,
which is significantly fainter in the $V$-band image, is a
saddle-point image (see \S\ref{sec:model}) which is often demagnified
by microlensing \citep{schechter02} makes the microlensing
interpretation plausible. Additional imaging observations in the same
and different bands will help to distinguish these possibilities.  

We note that the case for gravitational lensing is already very strong
from its characteristic configuration of image components, the
similarity of their colors, and the presence of a bright lensing
galaxy. However, the ultimate confirmation of its lensing nature will
require the spectroscopy of all image components to show they have
similar spectral features. 

\subsection{Optical Spectroscopy}
\label{sec:spec}

\begin{figure}
\begin{center}
 \includegraphics[width=0.9\hsize]{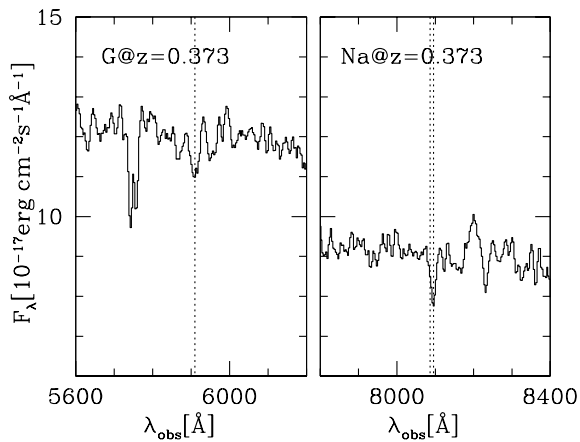}
\end{center}
\caption{G-band ({\it left}) and Na ({\it right}) absorption lines
  of the lensing galaxy at $z=0.373$ from the spectrum of
  SDSS~J1330+1810 obtained with the DIS at the ARC3.5m telescope. 
 The feature at 5750{\AA} is an MgII doublet at $z=1.054$ (see also
 Figure \ref{fig:spec1330}).
\label{fig:specna}}
\end{figure}

\begin{figure}
\begin{center}
 \includegraphics[width=0.95\hsize]{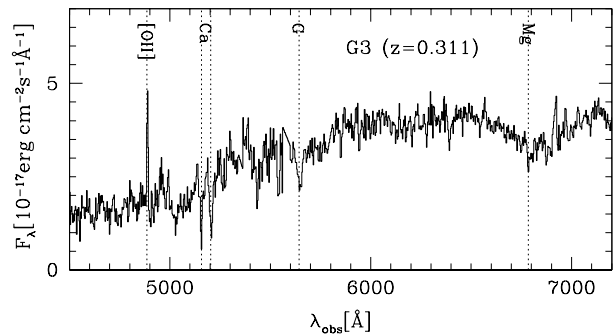}
\end{center}
\caption{The spectrum of galaxy G3 (See Figure \ref{fig:fc1330}) taken
  with the DIS at the ARC3.5m telescope. 
\label{fig:specg3}}
\end{figure}

We conducted spectroscopy of the system with the Dual Imaging
Spectrograph (DIS) at the ARC3.5m telescope on 2008 June 6. We used a
$1\farcs5$ slit and the gratings of B400 (dispersion of 1.8{\AA}) in
the blue channel and R300 (2.3{\AA}) in the red channel. The slit was
aligned to observe nearby galaxy G3 (see 
Figure \ref{fig:fc1330}) and SDSS~J1330+1810 simultaneously. The
spectral resolution is $R\sim 500$. We obtained the spectrum of
the Northern-half of SDSS~J1330+1810 (around image C and D) where the
relative contribution of galaxy G to the flux is much stronger;
however, because of poor seeing ($\sim 1\farcs2$), we could not
separate components C, D, and G. The exposure time was 2400~sec. The
data were reduced using standard IRAF tasks. 

This spectrum shows a weak absorption line at $\sim 5900${\AA} and
moderate absorption line at $\sim 8100${\AA} (Figure
\ref{fig:specna}), in excellent agreement with those of the G-band
4304{\AA} absorption and Na 5889/5896{\AA} absorption doublet
redshifted to $z=0.373$, the lens redshift inferred from absorption
lines in the SDSS spectrum (see Figure \ref{fig:spec1330}). Since
these absorptions are weak, they were not seen in the SDSS
spectrum. We note that the DIS spectrum exhibits clear emission lines
of quasars, which should originate both from images C/D and from
scattered fluxes from images A/B.  

The redshift of galaxy G3 is $z=0.311$ (Figure \ref{fig:specg3}),
which is quite close to the redshift of galaxy G2, $z=0.3126$. 
The two galaxies have similar spectra, with significant [OII] emission
of the same rest-frame equivalent width. The 4000{\AA} break of G3 is
weaker than that of G2, which implies that star formation happened
more recently in G3.  

\section{Environment of the Lens Galaxy}
\label{sec:env}

\begin{figure}
\begin{center}
 \includegraphics[width=0.9\hsize]{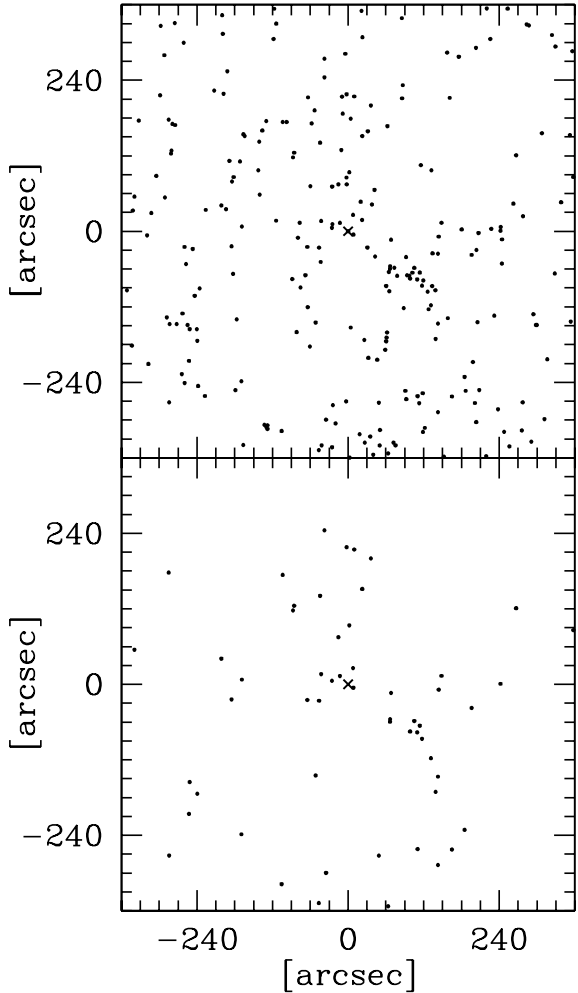}
\end{center}
\caption{{\it Upper:} The spatial distribution of all the SDSS
  galaxies brighter than $i=21$ in the vicinity of the lens at (0,
  0). North is up and East is left. The position of the lens system is
  indicated by a cross. {\it Lower:} The spatial distribution of
  galaxies whose photometric redshifts are consistent with $z=0.31$. 
    \label{fig:env}}
\end{figure}

The wide-field image suggests a possible group of galaxies
located near the lens system (see Section \ref{sec:sdss}). The high 
incidence of groups near strong lens systems has been noted before
\citep{fassnacht02,momcheva06,williams06,auger07,cabanac07,shin08,treu08} 
and is theoretically expected \citep{keeton01,oguri05}. 
Although the redshift of the possible group, $z=0.31$, differs from
the lens redshift $z_l=0.373$, a foreground group affects the lens
potential in a similar way as does a group at the same
redshift.\footnote{From the discussion of \citet{keeton03}, it is 
estimated that a weak foreground external shear $\gamma$ at
$z=0.31$ is equivalent to an external shear of $\approx 0.8\gamma$ at
$z=0.373$.} In this section, we study the distribution and 
properties of galaxies in the field from the SDSS.   

First we extract locations and brightnesses of galaxies in the
$12'\times12'$ field centred on the lens system from the SDSS data
release 6 \citep[DR6;][]{adelman08}. We adopt the \citet{petrosin76}
magnitudes in what follows. We restrict our analysis to
galaxies brighter than $i=21$, where star-galaxy separation is
reliable. To study the distribution at $z\sim 0.31$, we adopt
photometric redshift measurement in the SDSS databases
\citep{csabai03,adelman07}.  Specifically, we select galaxies whose
photometric redshifts are consistent with $z=0.31$ within their quoted 
errors. Therefore our study here is not necessarily restricted to red
elliptical galaxies but includes blue galaxies as well. Galaxies with
large photometric redshift errors, $\Delta z>0.2$, are excluded
from our analysis. 

We show the angular distributions of all galaxies and galaxies at
$z\sim 0.31$ in Figure \ref{fig:env}. There is a clear concentration
of galaxies to the southwest of the lens. The structure is more
pronounced after applying a cut by the photometric redshift. We note
that G1 is the brightest among galaxies at $z\sim 0.31$ selected in
this way. Together with the agreement of the spectroscopic redshifts
of G2 and G3 (Sections \ref{sec:sdss} and \ref{sec:spec}), we conclude
that there is a foreground group of galaxies at $z\sim0.31$ with its
centre $\sim 120''$ southwest of the lens system (corresponding to
transverse physical distances of 390$h^{-1}$kpc at $z=0.31$ and
440$h^{-1}$kpc at $z=0.373$), centred around G1 and G2.  It is worth
noting that a concentration of galaxies can also be seen around G3,
which suggests that there might be a sub-clump of the group around G3.  

There is a possibility that a concentration of galaxies at the lens
redshift $z=0.373$ exists in additional to the group at $z=0.31$. 
However, most of galaxies examined here are rather faint and have
large errors on the photometric redshifts, which prevent us to
distinguish structures at $z=0.31$ from $z=0.373$. Additional imaging
and spectroscopic follow-up observations are necessary to explore this
issue further. 

\section{Mass Modelling}
\label{sec:model}

\begin{table*}
 \caption{Best-fit mass models of SDSS J1330+1810. The column `Flux'
   shows whether flux ratios are included as constraints or not. The
   parameter $\theta_{\rm Ein}$, $e$, and $\gamma$ denote the Einstein
   radius, ellipticity, and external shear, respectively. The position
   angle of ellipticity and shear, $\theta_e$ and $\theta_\gamma$, are
   measured East of North. $\chi^2_{\rm pos}$, $\chi^2_{\rm gal}$, and
   $\chi^2_{\rm flux}$, indicate chi-square values from image
   positions, lens galaxy positions, and flux ratios. The total
   chi-square $\chi^2$ is the sum of these three. Also shown are time
   delays between images predicted by the best-fit models, adopting
   the lens redshift $z_l=0.373$. 
\label{table:model}} 
 \begin{tabular}{@{}cccccccccccccc}
  \hline
   Model & Flux  & $\chi^2$/dof  & $\chi^2_{\rm pos}$ &
   $\chi^2_{\rm gal}$ & $\chi^2_{\rm flux}$ &
   $R_{\rm Ein}$ & $e$ & $\theta_e$[deg] &
   $\gamma$ & $\theta_\gamma$[deg] & $\Delta t_{AB}$[day] & $\Delta t_{AC}$[day] &
   $\Delta t_{AD}$[day] \\ 
    \hline
SIE  & No & 2.10/3 & 2.10 & 0.01 & $\cdots$ & $0\farcs97$ & 0.31 & 34
   & $\cdots$ & $\cdots$ & $-0.15$ & 6.04 & $-11.68$\\
SIE  & Yes& 7.28/6 & 2.80 & 0.01 & 4.47 & $0\farcs97$ & 0.32 & 34 &
   $\cdots$ & $\cdots$ & $-0.18$ & 5.91 & $-11.83$\\
SIEx & No & 0.33/1 & 0.13 & 0.21 & $\cdots$ & $0\farcs97$ & 0.39 & 25
   & 0.05 & $-88$ & $-0.20$ & 6.76 & $-12.17$\\
SIEx & Yes& 1.14/4 & 0.76 & 0.13 & 0.25 & $0\farcs99$ & 0.56 & 25 &
   0.11 & $-75$ & $-0.31$ & 10.28 & $-12.69$\\ \hline
 \end{tabular}
\end{table*}

We constrain the lens potential of this system using the observed
image positions and flux ratios. The relative positions of the quasar
images and galaxies and their errors are adopted from Table
\ref{table:sdss1330}.  Flux ratios are estimated from the median of
the measurements in the five follow-up images (see Figure
\ref{fig:fratio}). Note that the flux ratios agree well with those of
the $H$-band image for which the effect of dust extinction is small.
Considering the scatter, we assume a conservative
error on the magnitude difference between quasar images,
$\sigma(\Delta m)=0.2$. We adopt two standard mass models that are
widely used in modelling strong lens systems; a Singular Isothermal
Ellipsoid (SIE) and an SIE plus external shear perturbation
(SIEx). Since it is of interest whether the models can explain the
observed flux ratios or not, for each model we consider two cases; one
includes flux ratios as observational constraints and the other only
uses relative positions as constraints. The optimisation of model
parameters is performed using a software package called {\it glafic}
(M. Oguri, in preparation). 

\begin{figure}
\begin{center}
 \includegraphics[width=0.95\hsize]{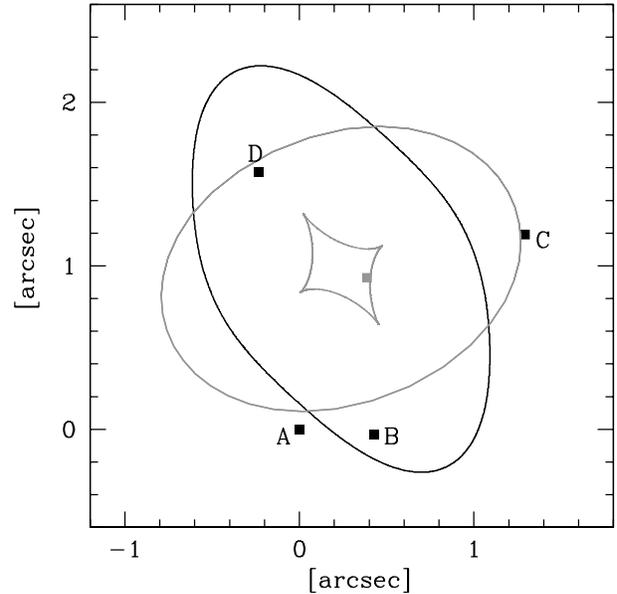}
\end{center}
\caption{The best-fit model (SIEx, flux ratios are included as
  constraints) is shown. See Table \ref{table:model} for best-fit
  parameters. The black curve and squares indicate the critical curve
  of the best-fit model and the predicted image positions. Observed
  image positions are within the symbols. Grey curves and the square
  are the corresponding caustics and source position.  
\label{fig:crit1330}}
\end{figure}

Best-fit $\chi^2$ and model parameters are summarised in Table
\ref{table:model}. We find that all four models considered here 
produce reasonably good fits to the data ($\chi^2/{\rm dof}\la 1$).
The models successfully fit the flux ratios, therefore this lens
system does not exhibit any sign of flux ratio anomaly. Moreover, the
flux ratios predicted by models with no flux ratio constraints broadly
agree with observed flux ratios. The successful mass modeling further
supports the lensing nature of this system. We show the critical curve
and caustics for the best-fit model (SIEx with flux ratios as
constraints) in Figure \ref{fig:crit1330}, which indicates that images
B and D are saddle-point images and images A and C are minima. Given
the best-fit mass models, we can predict time delays between the
images. Assuming that the lensing galaxy is located at $z_l=0.373$,
for each  best-fit model we compute  time delays between image A and
the other images, which are listed in Table \ref{table:model}. Time
delays are rather model dependent, particularly for $\Delta t_{AB}$
and $\Delta t_{AC}$, for which the maximum fractional differences
between different models are $\ga 70\%$. This is because shorter time
delays are more easily affected by perturbations in the primary lens
potential \citep{oguri07a}.   

\begin{figure}
\begin{center}
 \includegraphics[width=0.95\hsize]{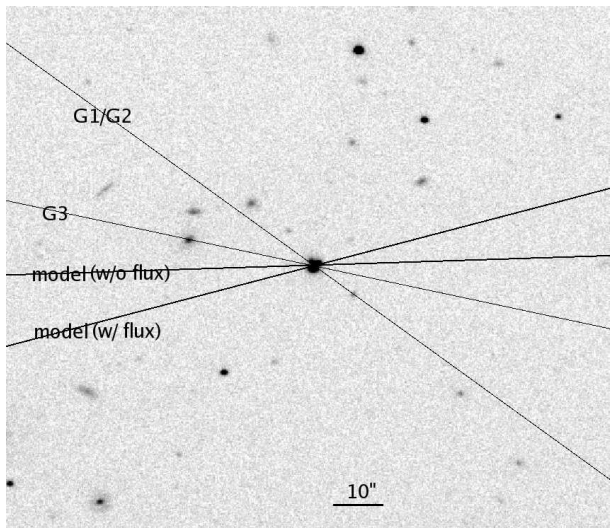}
\end{center}
\caption{The UH88 $V$-band image around the lens system. North is up
  and East is left. Shear directions in our best-fit models
  ($\theta_\gamma=-88^\circ$ for the model without flux constraints,
  and $-75^\circ$ for the model with flux constraints) are shown by
  thick solid lines. Directions to a nearby bright galaxy (G3) and the
  centre of the group (G1/G2) are indicated by thin solid lines.
\label{fig:uh_shear}}
\end{figure}

\begin{figure}
\begin{center}
 \includegraphics[width=0.95\hsize]{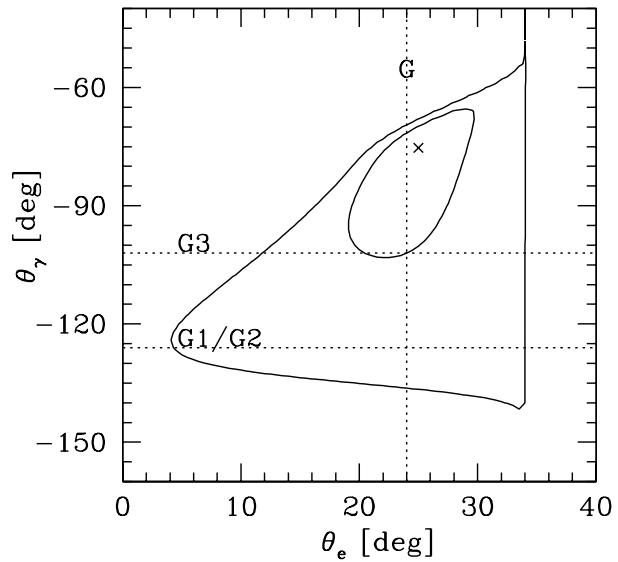}
\end{center}
\caption{1$\sigma$ and 2$\sigma$ contours ($\Delta\chi^2=2.3$ and
  $6.2$ for 2 degree of freedom) of the position angles of
  the ellipticity and external shear for the SIEx model. Note that the
  other model parameters are marginalised over. Flux ratios
  are included as constraints. The cross indicates the best-fit values 
  (see Table \ref{table:model}). The vertical dotted line indicates
  the observed orientation of the lensing galaxy. Horizontal dotted
  lines are expected values (see also Figure~\ref{fig:uh_shear}) when
  the external perturbations are caused by the centre of the group
  (G1/G2) or the nearby bright galaxy (G3). 
\label{fig:eg}}
\end{figure}

The large ellipticities of $0.31-0.56$ in the best-fit models are
broadly consistent with the observed shape of the lensing galaxy (see
Section \ref{sec:image}). We find that the best-fit position angle of
the galaxy, $\theta_e$, is different between models with and without
external shear: $\theta_e\sim 34^\circ$ when the external shear is not
included, whereas $\theta_e\sim 25^\circ$ for models including
external shear. The observed position angle of galaxy G,
$\theta_e= 24^\circ\pm2^\circ$, agrees very well with the latter,
suggesting the 
non-negligible effect of external shear on the lens potential, as in
other quadruple lens systems \citep{keeton97}. However, there is no
obvious perturber along the shear direction, $\theta_\gamma=-88^\circ$
or $-75^\circ$ (see Figure \ref{fig:uh_shear}). Thus in Figure
\ref{fig:eg} we present likelihood contours in the
$\theta_e$-$\theta_\gamma$ plane. Specifically, for each ($\theta_e$, 
$\theta_\gamma$) we perform $\chi^2$ minimisations and use
$\Delta\chi^2$ to draw contours at the 1$\sigma$ and 2$\sigma$
confidence levels. We find that the shear direction to galaxy G3 is
consistent with the observations at the 1$\sigma$ level. The resultant
ellipticity position angle is $20^\circ\la \theta_e\la24^\circ$ and is
also consistent with the observed orientation of the lensing
galaxy. On the other hand, the shear direction to galaxies G1 and G2
fails to fit the data at the $\sim 2\sigma$ level. Therefore we
conclude that nearby galaxy G3, rather than the group of galaxies
$\sim 120''$ southwest of the lens, is likely to be the main perturber
to the lens system. 

Although the shear amplitudes of $0.05-0.1$ are slightly larger than
would be produced by a single galaxy $\sim 30''$ away from the lens, 
the existence of the group suggests that dark matter sufficient to
produce the shear amplitudes can be associated with galaxy G3. This
idea is supported by an apparent concentration of galaxies at $z\sim
0.31$ around G3 (see Figure \ref{fig:env}).

Given the best-fit Einstein radius $R_{\rm Ein}=0\farcs97$, we can
compute magnitudes of the lensing galaxy expected from the
Faber-Jackson relation \citep{faber76}, adopting the model of
\citet{rusin03}. From the lens redshift $z_l=0.373$, the apparent
magnitudes are predicted to be $V=20.7$, $J=17.9$, $H=16.7$ and
$K_s=16.1$, which are $0.7-1.2$~mag fainter than the observed total
magnitudes of the lensing galaxy. Using the same model, we find that
the observed magnitudes correspond to the lens redshift of $z_l\sim
0.22$. The discrepancy is substantially larger than the observed
1$\sigma$ scatter in the Faber-Jackson relation ($\sim 0.5$~mag). While it is
not clear what causes the difference, a possible explanation is an
interaction with nearby structure which can strip the outer part of the
galaxy mass and reduce the Einstein radius without changing the
overall luminosity of the galaxy \citep[e.g.,][]{treu08}. However, there
is no obvious companion galaxy at  $z\sim 0.373$ around galaxy G, and no
obvious signature of interaction in the galaxy images.  Again,
extensive spectroscopy of galaxies in the field to study structure
around the lens redshift might help to resolve this issue.  

\section{Summary}
\label{sec:sum}

We have presented the discovery of a new four-image lensed quasar
SDSS J1330+1810 ($z_s=1.393$). This source was selected as a lens
candidate by the SQLS due to its extended morphology. Our 
observations in optical and near-infrared indicate that it is a
typical fold-type quadruple lens, with a maximum separation between
images of $1\farcs76$.  From the spectrum we measured a lens
redshift of $z_l=0.373$. Standard simple elliptical mass models fit
the data well, including the flux ratios, implying no evidence for
substructure. The mass modelling suggests an important contribution of  
the external shear, probably from a nearby bright galaxy. There is
also a foreground group of galaxies whose centre is $\sim120''$ from
the lens. The lens galaxy is $\sim 1$~mag brighter than predicted by
mass modelling. 

Thus far the SQLS has discovered 31 lensed quasars including SDSS
J1330+1810\footnote{The current status of the survey is summarised 
at http://www-utap.phys.s.u-tokyo.ac.jp/\~{}sdss/sqls/}, of which 5 are
quadruple lenses \citep[e.g.,][]{inada03,kayo07}. The low fraction of
quadruple lenses may imply that the faint end slope of the quasar
optical luminosity function is shallow, although it is marginally
consistent with standard theoretical expectations \citep{oguri07b}. 
The SQLS has completed $\sim 2/3$ of its survey, implying that a few
more quadruple lenses will be discovered by the completion of the survey.

\section*{Acknowledgments}

We thank Paul Schechter for his help of the Magellan observation.
This work was supported in part by Department of Energy contract
DE-AC02-76SF00515 and RIKEN DRI Research Grant.
J.~A.~B.  acknowledges support from US NSF grant AST-0206010.
I.~K. acknowledges support from the JSPS Research Fellowships 
for Young Scientists and Grant-in-Aid for Scientific Research on
Priority Areas No. 467. M.~A.~S. acknowledges support of NSF grant
AST-0707266. 

Funding for the SDSS and SDSS-II has been provided by the Alfred
P. Sloan Foundation, the Participating Institutions, the National
Science Foundation, the U.S. Department of Energy, the National
Aeronautics and Space Administration, the Japanese Monbukagakusho, the
Max Planck Society, and the Higher Education Funding Council for
England. The SDSS Web Site is http://www.sdss.org/.

The SDSS is managed by the Astrophysical Research Consortium for the
Participating Institutions. The Participating Institutions are the
American Museum of Natural History, Astrophysical Institute Potsdam,
University of Basel, University of Cambridge, Case Western Reserve
University, University of Chicago, Drexel University, Fermilab, the
Institute for Advanced Study, the Japan Participation Group, Johns
Hopkins University, the Joint Institute for Nuclear Astrophysics, the
Kavli Institute for Particle Astrophysics and Cosmology, the Korean
Scientist Group, the Chinese Academy of Sciences (LAMOST), Los Alamos
National Laboratory, the Max-Planck-Institute for Astronomy (MPIA),
the Max-Planck-Institute for Astrophysics (MPA), New Mexico State
University, Ohio State University, University of Pittsburgh,
University of Portsmouth, Princeton University, the United States
Naval Observatory, and the University of Washington.
 
This publication makes use of data products from the Two Micron All
Sky Survey, which is a joint project of the University of
Massachusetts and the Infrared Processing and Analysis
Center/California Institute of Technology, funded by the National
Aeronautics and Space Administration and the National Science
Foundation.



\label{lastpage}

\end{document}